\documentclass[aps,prc,twocolumn,showpreprintnumbers,floatfix,nofootinbib]{revtex4-2}
\usepackage{amsmath,amssymb,graphicx,bm}
\usepackage{bbm}
\usepackage{color}
\usepackage[normalem]{ulem}
\usepackage{slashed}
\usepackage[colorlinks,allcolors=blue]{hyperref}
\usepackage{orcidlink}

\allowdisplaybreaks[4]
\newcommand{\vect}[1]{\bm{\mathrm{#1}}}
\newcommand{\pv}{\vect{p}}
\newcommand{\qv}{\vect{q}}
\newcommand{\sigmav}{\vect{\sigma}}

\newcommand{\nubar}{\overline{\nu}}

\newcommand{\tmin}{\text{min}}
\newcommand{\tmax}{\text{max}}
\newcommand{\emis}{\text{em}}
\newcommand{\abso}{\text{ab}}

\newcommand{\Sign}{\mathbb{S}}
\newcommand{\fm}{\rm{fm}}
\newcommand{\CC}{\text{CC}}
\newcommand{\NC}{\text{NC}}
\newcommand{\rnabla}{\overset{\raisebox{-0.25ex}{$\scriptscriptstyle\rightarrow$}}{\nabla}}
\newcommand{\lnabla}{\overset{\raisebox{-0.25ex}{$\scriptscriptstyle\leftarrow$}}{\nabla}}
\newcommand{\lrnabla}{\overset{\raisebox{-0.25ex}{$\scriptscriptstyle\leftrightarrow$}}{\nabla}}

\begin{document}
\title{Weak magnetism effects on neutrino-nucleon interactions in hot nuclear matter within nonrelativistic Skyrme RPA}
\author{Mingya Duan\orcidlink{0000-0001-5487-4537}{}}
\affiliation{Independent Researcher, Yunnan, China}
\author{Michael Urban\orcidlink{0000-0002-3328-0091}{}}
\email{michael.urban@ijclab.in2p3.fr}
\affiliation{Universit\'e Paris-Saclay, CNRS/IN2P3, IJCLab, 91405 Orsay, France}

\begin{abstract}
Neutrino production, absorption, and scattering are important in
explaining neutrino-driven core-collapse supernovae. At high densities
or temperatures, relativistic effects, in particular weak magnetism
(WM), may become increasingly important. Therefore, we extend the
nonrelativistic random-phase approximation (RPA) framework for the
study of neutrino-nucleon interactions in hot nuclear matter by
incorporating the WM term into the nucleon weak current. In
calculations with our recent Skyrme functional Sky3s, we confirm that
WM increases interaction rates for the most important charged-current
neutrino processes in supernovae and reduces those of
antineutrinos. However, this is not the case for the antineutrino
production via $\beta$ decay, which is also increased. In our
calculations, WM can almost double the neutrino emissivities of
neutron-star matter by direct Urca processes at high density. We also
find that WM corrections to the scattering rates are increasingly
important with increasing scattering angle and energy transfer,
leading to a significant difference between double-differential
neutrino and antineutrino scattering rates. Our formulas can help
improve neutrino rates needed in supernova simulations.
\end{abstract}

\maketitle

\section{Introduction}\label{sec:introduction}
At the end stage of the evolution of a massive star ($M \gtrsim 8
M_{\odot}$, $M_{\odot}$ is the solar mass), a core-collapse supernova
occurs \cite{Cerda-DuranElias-Rosa}. As the core collapses, the
density and temperature of the matter increase dramatically, a large
number of protons are transformed into neutrons, and neutrinos of all
flavors are produced. Neutrinos interact frequently with nuclear
matter before they escape from the star, namely, they can be absorbed
and scattered by nucleons. Numerous recent studies have concentrated
on various aspects of improving the treatment of neutrino-nuclear
matter interactions, such as taking into account relativistic
kinematics
\cite{Roberts2017,Fischer2020a,Guo2020,Sugiura2022,Sokolowski2026},
considering the role of muons
\cite{Bollig2017,Guo2020,Fischer2020b,Sugiura2022,Loffredo2023,Gieg2025,Sokolowski2026},
employing chiral effective field theory (EFT)
\cite{Bacca2009,Bacca2012,Rrapaj2015,Guo2019,Vidana2022,Shin2024},
using the Brueckner-Hartree-Fock (BHF) formalism with the two-body
potential of chiral EFT \cite{Guo2024}, and so on.

Among the neutrino-nuclear matter interactions, neutrino
absorption reactions are the most important ones for a
successful neutrino-driven core-collapse supernova explosion. During
the shock stagnation stage before a successful supernova explosion,
neutrinos deposit energy behind the stalled bounce shock mainly
through neutrino absorption reactions and to a lesser extent neutrino
scattering reactions \cite{Janka2017b}, resulting in neutrino
heating.

Neutrino production and absorption can be related to the
charged-current response functions of nuclear matter
\cite{Hernandez1999,Duan2026}, while neutrino scattering can be
related to the neutral-current ones \cite{Iwamoto1982}. The response
functions depend on the nuclear interaction. Skyrme interactions are
easy to use for computing response functions. Recently, we constructed
extended Skyrme functionals Sky3s and Sky4s \cite{Duan2024,Duan2025}
in order to revisit the problem of neutrino rates for the use in
astrophysical simulations. By using effective masses, equation of
state (EoS), and spin Landau parameters from microscopic BHF
calculations in the fit of these functionals, we avoid some
pathological or unrealistic behaviors which are present in SLy4
\cite{Chabanat1998} and other common Skyrme interactions
\cite{Duan2023,Duan2025}.
 
Some previous literature considers only the leading-order terms of the
nucleon weak currents, i.e., the vector (V) and axial-vector (A)
currents, which are the historically known dominant contributions in
nuclear $\beta$ decay \cite{BjorkenDrell}. In some references
\cite{Vogel1984,Burrows2000,Leinson2002,Leinson2002b,Horowitz2002,Horowitz2003,Burrows2006,Fischer2020a,Fischer2020b,Guo2020},
also subleading-order corrections are considered. Among them, the weak
magnetism (WM) term is of particular interest, since it acts
differently for neutrinos and antineutrinos
\cite{Horowitz2002}. However, to our knowledge, the WM term has not
been incorporated into neutrino-rate calculations with nonrelativistic
Skyrme functionals. Besides, most previous studies of the WM
corrections to the charged current considered only some of the
neutrino processes in neutrino-driven core-collapse supernovae, such
as $e^{-} + p \leftrightarrow \nu_e + n$ and $e^{+} + n
\leftrightarrow \nubar_e + p$, but neglected others, such as $n
\leftrightarrow p+e^{-} + \nubar_e$ and $p\leftrightarrow
n+e^{+}+\nu_e$. In other words, the last two processes (and the
corresponding ones with muons) have not been completely studied with
WM corrections.

In this work, we study the effects of WM on the neutrino interactions
in hot nuclear matter using the nonrelativistic Skyrme functional
Sky3s \cite{Duan2025} in the full random phase approximation (RPA). We
first incorporate the WM term into the nucleon weak currents and
detail the modified formulas for neutrino production, absorption, and
scattering rates in Sec. \ref{sec:neutrino-nucleon interactions in hot
  nuclear matter}. Then, we compute the neutrino opacities and direct
Urca neutrino emissivities with the WM corrections, and compare them
to those obtained without WM in
Sec. \ref{sec:production-absorption}. Also, we study the effects of WM
corrections on neutrino scattering in
Sec. \ref{sec:scattering}. Finally, we summarize and conclude in
Sec. \ref{sec:conclusion}.

\section{Neutrino-nucleon interactions in hot nuclear matter} 
\label{sec:neutrino-nucleon interactions in hot nuclear matter}
Recently, Refs. \cite{Duan2023,Duan2026} recalled the derivation of
the relation between the neutrino scattering, production and
absorption rates and the nuclear density and spin-density correlation
functions in a non-relativistic framework. In this section, we
incorporate in addition the WM term into the nucleon weak currents,
which was not included in Refs. \cite{Duan2023,Duan2026}. Throughout
this paper, we use units with $\hbar = c = k_B = 1$, where $\hbar$,
$c$, and $k_B$ are the reduced Planck constant, the speed of light,
and the Boltzmann constant, respectively.

We are interested in charged-current (CC) neutrino production
and absorption processes of the types
\begin{gather}
  b\leftrightarrow l + \nu + b'\,, \label{eq:cc-process13}\\
  b + l \leftrightarrow b' + \nu\,, \label{eq:cc-process22}
\end{gather}
as well as neutrino scattering via the neutral current (NC),
\begin{equation}\label{eq:nc-process}
\nu + b \rightarrow \nu + b\,,
\end{equation}
where $b, b'$ denote baryons (nucleons) $n$ or $p$, $l$ stands
generically for a charged lepton $e^-$, $e^+$, $\mu^-$, or $\mu^+$,
and $\nu$ for a neutrino $\nu_f$ or antineutrino $\bar{\nu}_f$ of
flavor $f$ ($e$, $\mu$, or $\tau$), respecting charge, lepton number
and flavor conservation.

Let us use the same notations as in Refs. \cite{Duan2023,Duan2026} and
write the weak lepton-nucleon interaction Lagrangian density as
\cite{Reddy1998,Iwamoto1982}
\begin{equation}\label{eq:lagrangian density}
\mathcal{L}_{I} = \frac{G}{\sqrt{2}}
(l^\dagger_{\CC\,\lambda} j_{\CC}^{\lambda}+l_{\CC\,\lambda} j_{\CC}^{\dagger\, \lambda})
    + \frac{G_F}{\sqrt{2}}l_{\NC\,\lambda} j_{\NC}^\lambda,
\end{equation}
where $G_F \approx 8.975 \times 10^{-5}$ MeV fm$^3$ is the Fermi weak
interaction constant, $\cos\theta_c = 0.974$ is the cosine of the
Cabibbo angle, and $G=G_F \cos \theta_c$.  The relevant terms of the
lepton weak currents have the form
\begin{align}\label{eq:lepton weak current}
  l_{\CC}^{\lambda} &=\sum_{f=e,\mu}\hat{\bar{\psi}}_{\nu_f} \gamma^{\lambda}
    (1-\gamma_5) \hat{\psi}_{f},\\
  l_{\NC}^{\lambda} &=\sum_{f=e,\mu,\tau}\hat{\bar{\psi}}_{\nu_f} \gamma^{\lambda}
    (1-\gamma_5) \hat{\psi}_{\nu_f}.
\end{align}
The nucleon weak currents have additional terms
from the nucleon anomalous magnetic moments and are given by
\begin{align}
  j^{\lambda}_{\CC} & = \hat{\bar{\psi}}^{}_{p} \left[ \gamma^{\lambda}
    (g^{}_{V} - g^{}_{A} \gamma^{5}) + g^{}_M \frac{i \sigma^{\lambda \alpha}
      q_{\alpha}}{2m_N} \right] \hat{\psi}^{}_{n},
  \label{eq:nucleon weak current-WM}\\
  j^{\lambda}_{\NC} & = \sum_{b = n,p} \frac{1}{2} \hat{\bar{\psi}}^{}_{b}
    \left[\gamma^{\lambda} (C^{}_{V,b} - C^{}_{A,b} \gamma^{5})+ C^{}_{M,b}
      \frac{i \sigma^{\lambda \alpha} q_{\alpha}}{2m_N}\right] \hat{\psi}^{}_{b},
  \label{eq:nucleon weak neutral current-WM}
\end{align}
where $g^{}_{V}=1$, $g^{}_{A}=1.23$, $C_{V,n}^{}=-1$,
$C_{A,n}^{}=-1.23$, $C_{V,p}^{}=0.08$, and $C_{A,p}^{}=1.23$
\cite{Reddy1998}. The additional couplings are $g^{}_M=3.706$,
$C_{M,n}^{}=-1.926$, and $C_{M,p}^{}=2.038$ \cite{Horowitz2002}, and
$m_N$ is the nucleon mass.
\footnote{In the literature, different conventions for lepton and
nucleon weak currents exist, which differ by constant factors that can
be instead included in the prefactors of Eq.
\eqref{eq:lagrangian density}.}

When making the nonrelativistic reduction of the field operators, we
now have to consider both the leading and subleading orders in
$\pv/m_N$ (where $\pv$ is the nucleon momentum) instead of only the
leading order as done in Refs. \cite{Duan2023,Duan2026}, due to the
introduction of the WM term. Then, the nonrelativistic
charged current takes the form
\begin{align}
  j^{0}_{\CC} =& \hat{\varphi}_{p}^{\dagger}
  \Big( g^{}_V + ig^{}_A \frac{\sigmav\!\cdot\!\lrnabla}{2m_N}\Big)
  \hat{\varphi}_n, \nonumber \\
  \vect{j}_{\CC} =& \hat{\varphi}_{p}^{\dagger} \Big(-g^{}_A \sigmav
    - ig^{}_V \frac{\lrnabla}{2m_N}
    + \tilde{g}^{}_M \frac{(\lnabla\!+\!\rnabla)\!\times \! \sigmav}{2m_N}
    \Big) \hat{\varphi}_n,\label{eq:jCC}
\end{align}
where $\hat{\varphi}_b$ are the corresponding nonrelativistic field
operators, $\sigmav$ are the Pauli matrices, $\tilde{g}^{}_M =
g^{}_V+g^{}_M$, and $\lrnabla = \rnabla-\lnabla$. Notice that, when
taking matrix elements of $j_{\CC}^\lambda$ between plane-wave states
with nucleon momenta $\pv_p$ and $\pv_n$, one gets $\lrnabla\to
i(\pv_n\!+\pv_p)$ and $\lnabla\!+\!\rnabla \to -i(\pv_p-\pv_n)$. Let
us keep only the leading-order terms and the magnetic one because it
has the large factor $\tilde{g}^{}_M=4.706$. Then $j_{\CC}^\lambda$
can be written as
\begin{equation}\label{eq:nucleon weak current-approx-WM}
  j^{\lambda}_{\CC} \approx \hat{\varphi}_{p}^{\dagger}\Big(g^{}_V,
    -g^{}_{A} \sigmav
    + \tilde{g}^{}_M \frac{(\lnabla\!+\!\rnabla)\!\times\! \sigmav}{2m_N}
    \Big)^\lambda \hat{\varphi}^{}_{n}.
\end{equation}

Similarly, we can get for the neutral-current case
\begin{align}\label{eq:nucleon weak neutral current-approx-WM}
  j^{\lambda}_{\NC} \approx \sum_{b=n,p} \frac{1}{2} \hat{\varphi}_{b}^{\dagger}
  \Big(C^{}_{V,b}, - C^{}_{A,b} \sigmav
    + \tilde{C}_{M,b} \frac{(\lnabla\!+\!\rnabla)\! \times\! \sigmav}{2m_N}
  \Big)^{\lambda} \hat{\varphi}^{}_{b},
\end{align}
where $\tilde{C}_{M,b} = C_{V,b}+C_{M,b}$.

Following the steps from Eq. (8) to Eq. (18) of Ref. \cite{Duan2026},
the differential rates $dR/d^3pd^3x$ of all processes
mentioned in Eqs. \eqref{eq:cc-process13} and
  \eqref{eq:cc-process22} can be generically written as
\begin{widetext}
\begin{align}\label{eq:differential rates for all processes}
 \frac{dR}{d^3p_{\nu} d^3 x}
 = & \frac{G^2}{(2 \pi)^4} \int_{q_{\tmin}}^{q_{\tmax}} dq
 \int_{\omega_{\tmin}}^{\omega_{\tmax}} d \omega  q
   \frac{E_l}{E_{\nu}} F_l F_{\nu}
   \Big\{ \Big(1 + \frac{p_l}{E_l} \cos\theta \Big) g_{V}^2
   S_{b'b}^{(S=0)}(q, \omega)
   +\Big(3- \frac{p_l}{E_l} \cos\theta \Big) g_A^2 S_{b'b}^{(SM=11)}(q,\omega)
 \nonumber \\  
 & + \Big(1+\frac{p_l}{E_l} \cos\theta
   +\Sign_l \Sign_{\nu} \frac{2p_l^2 E_{\nu}}{E_l q^2} \sin^2\theta \Big)
   g_A^2 \Big[S_{b'b}^{(SM=10)}(q,\omega)-S_{b'b}^{(SM=11)}(q,\omega)\Big]
 \nonumber \\
 & +2 \Big( q^2 - \frac{p_l}{E_l} q^2  \cos \theta
   - \Sign_l \Sign_{\nu} \frac{p_l^2 E_{\nu}}{E_l} \sin^2 \theta  \Big)
   \frac{\tilde{g}_M^2}{4m_N^2} S_{b'b}^{(SM=11)}(q,\omega)
 \nonumber \\
 & + 4 \Sign_p \Big[ \Sign_l  \frac{p_l^2}{E_l} - \Sign_{\nu}  E_{\nu}
   - \frac{p_l}{E_l} ( \Sign_l  E_l - \Sign_{\nu} E_{\nu}) \cos \theta \Big]
 \frac{g^{}_A\tilde{g}^{}_M}{2m_N}
   S_{b'b}^{(SM=11)}(q,\omega) \Big\}.
\end{align}
\end{widetext}
In this equation, $S_{b'b}^{(SM)}(q,\omega)$ are the charged-current
$b'b^{-1}$ response functions as defined in Ref. \cite{Duan2026} for
spin and spin projection $S$ and $M$, $\omega$ and $\qv$ are the
energy and momentum transferred from the leptons to the nuclear matter
(integrated within their kinematically allowed limits), $\pv_i$ and
$E_i$ denote, respectively, the momentum and energy of an incoming or
outgoing particle $i$, and $\theta$ is the angle between $\pv_l$ and
$\pv_\nu$. To account for all the different processes in a single
equation, the signs $\Sign_i$ indicate if a particle $i$ is incoming
$(\Sign_i=1)$ or outgoing ($\Sign_i=-1$), and $F_i$ denotes the
occupation number $n^i_{\pv_i}$ for incoming particles or the Pauli
blocking factor $1-n^i_{\pv_i}$ for outgoing ones, see Table I of
Ref. \cite{Duan2026}. Note that there is an additional sign $\Sign_p$
which was not needed in the case without WM contributions in
Ref. \cite{Duan2026}.

We can see that there is a $g^{}_A\tilde{g}^{}_M$ interference term in
addition to the $\tilde{g}_M^2$ term in the expression
\eqref{eq:differential rates for all processes}. This term appears
because the term $i\varepsilon_{ij\gamma \delta}
p_{\nu}^{\gamma}p_{l}^{\delta}$ that one gets when working out the
lepton part ($i,j$ being the indices of the nucleon charged-current
three-vectors) can now contribute due to the introduction of
WM. Equation \eqref{eq:differential rates for all processes} shows
that the WM contributions are related to the transverse spin response
function ($S=1$, $M=\pm 1$ channel). From the two additional terms
(the last two lines) in Eq. \eqref{eq:differential rates for all
  processes}, one can expect that the WM leads to corrections with
different signs for different processes. Using the charged-current RPA
response functions developed in Ref. \cite{Duan2026}, the differential
rates for all processes mentioned in
Eqs. \eqref{eq:cc-process13} and \eqref{eq:cc-process22}
can be computed.

Similarly, the double-differential (anti-)neutrino scattering rates
via the neutral-current interaction can be obtained by
\begin{widetext}
\begin{align}\label{eq:scattering rate-sum}
  \frac{dR}{d \cos\theta d E'_{\nu}} = & \frac{G_{F}^2}{8\pi} E_{\nu}^{\prime\,2}
  \sum_{bb'=n,p}\bigg\{ (1 + \cos\theta) C_{V,b} C_{V,b'}
  S_{bb,b'b'}^{(S=0)}(q,\omega)
  + (3-\cos\theta) C_{A,b} C_{A,b'} S_{bb,b'b'}^{(SM=11)}(q,\omega)
  \nonumber \\
 & +  \Big(1 + \cos\theta - \frac{2E_{\nu} E'_{\nu} \sin^2\theta}{q^2} \Big)
 C_{A,b}C_{A,b'}
 \Big[ S_{bb,b'b'}^{(SM=10)}(q,\omega) - S_{bb,b'b'}^{(SM=11)}(q,\omega) \Big]
 \nonumber \\
 & +2(q^2-q^2\cos\theta + E_{\nu} E'_{\nu} \sin^2\theta)
 \frac{\tilde{C}_{M,b}\tilde{C}_{M,b'}}{4m_N^2}
 S_{bb,b'b'}^{(SM=11)}(q,\omega)
 \nonumber \\
 &  \pm 4(E_{\nu} + E'_{\nu})(1 - \cos\theta)
     C_{A,b} \frac{\tilde{C}_{M,b'}}{2m_N} S_{bb,b'b'}^{(SM=11)}(q,\omega)
 \bigg\},
\end{align}
\end{widetext}
where $S_{bb,b'b'}^{(SM)}$ are the neutral-current response functions
defined in Ref. \cite{Duan2023} (where they were called
$S_{bb'}^{(SM)}$, but here we change the notation to avoid confusion
with the charged-current response functions). Notice that while at the
Hartree-Fock level there are only $nn^{-1}$ and $pp^{-1}$ responses
$S_{nn,nn}$ and $S_{pp,pp}$, the residual interaction in the RPA mixes
$nn^{-1}$ and $pp^{-1}$ excitations, giving rise to $S_{nn,pp}$ and
$S_{pp,nn}$. The initial and final (anti-)neutrino momenta are denoted
$\pv_\nu$ and $\pv'_\nu$, respectively, and $E_\nu$ and $E'_\nu$ are
the corresponding energies, $\theta$ is the scattering angle, and $\qv
= \pv_\nu-\pv'_\nu$ and $\omega = E_\nu-E'_\nu$ are the momentum and
energy transferred from the neutrino to the nuclear matter.

In the last line of Eq. \eqref{eq:scattering rate-sum}, ``$+$'' is for
neutrinos and ``$-$'' is for antineutrinos. This difference in sign is
an important effect of the WM, since it can lead to different fluxes
of neutrinos and antineutrinos and thus affect the evolution of the
proton fraction during the supernova \cite{Horowitz2002}.

\begin{figure*}[tb!]
\begin{center}
\includegraphics[scale=0.6]{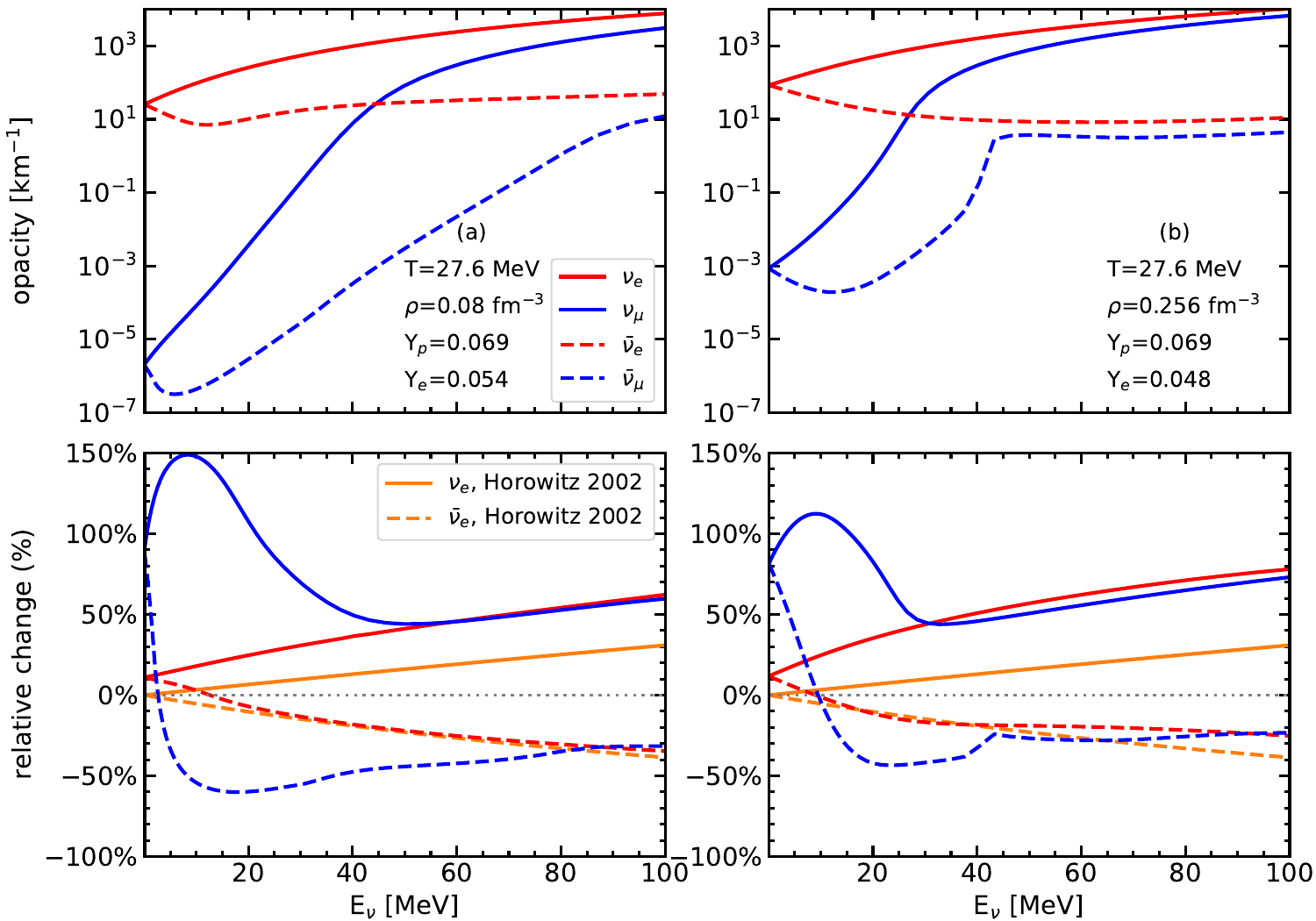} 
\caption{Top panels: (anti)neutrino opacities with WM corrections as
  functions of the neutrino energy $E_{\nu}$, computed within RPA
  using the Sky3s functional, for conditions (a) and (b). The red and
  blue solid lines represent the results for electron and muon
  neutrinos, while the red and blue dashed lines represent those for
  electron and muon antineutrinos, respectively. Bottom panels:
  corresponding relative changes compared to the opacities obtained
  without WM. For comparison, the orange solid and dashed lines show
  the relative changes of neutrino and antineutrino absorption given
  by Horowitz in Ref. \cite{Horowitz2002}.}
\label{fig:opacity-relative-change}
\end{center}
\end{figure*}
\begin{figure}[tb!]
\begin{center}
\includegraphics[scale=0.6]{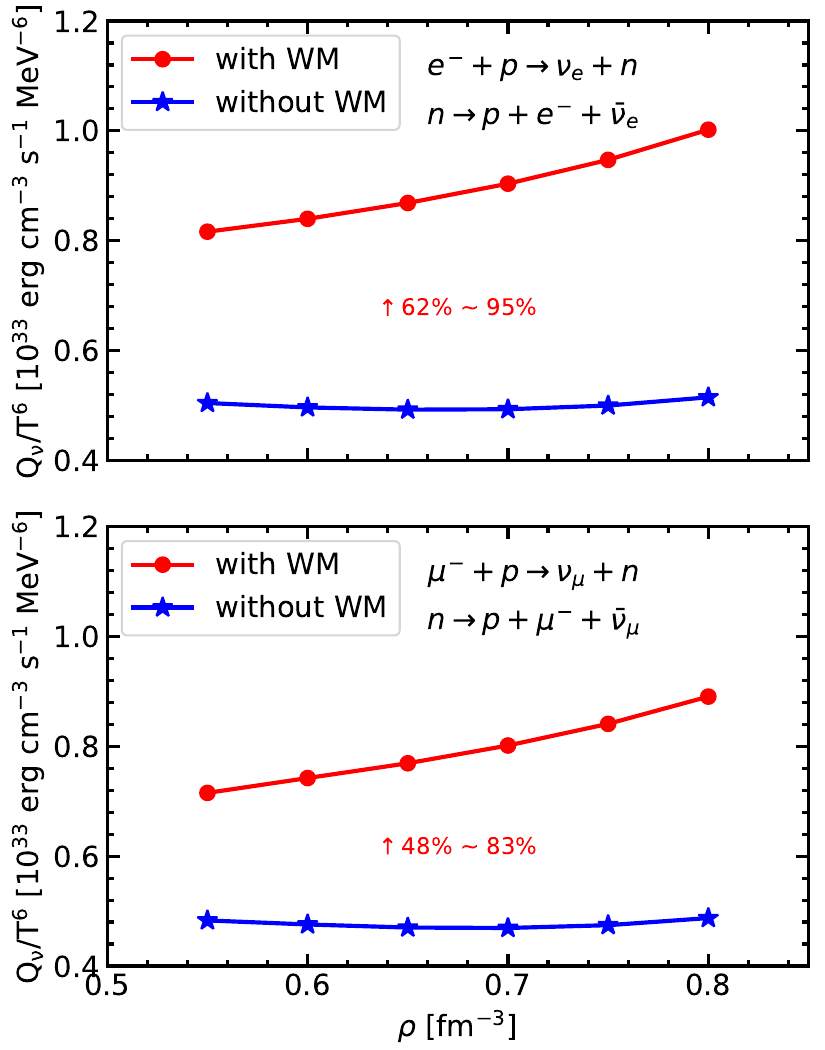} 
\caption{Direct Urca neutrino emissivities (divided by $T^6$, with $T$
  in MeV) of neutron-star matter with (red) and without (blue) weak
  magnetism for $npe\mu$ matter in $\beta$ equilibrium as functions of
  the baryon number density. Top: electron processes; bottom: muon
  processes.}
\label{fig:direct-Urca}
\end{center}
\end{figure}
\begin{figure*}[tb!]
\begin{center}
\includegraphics[scale=0.6]{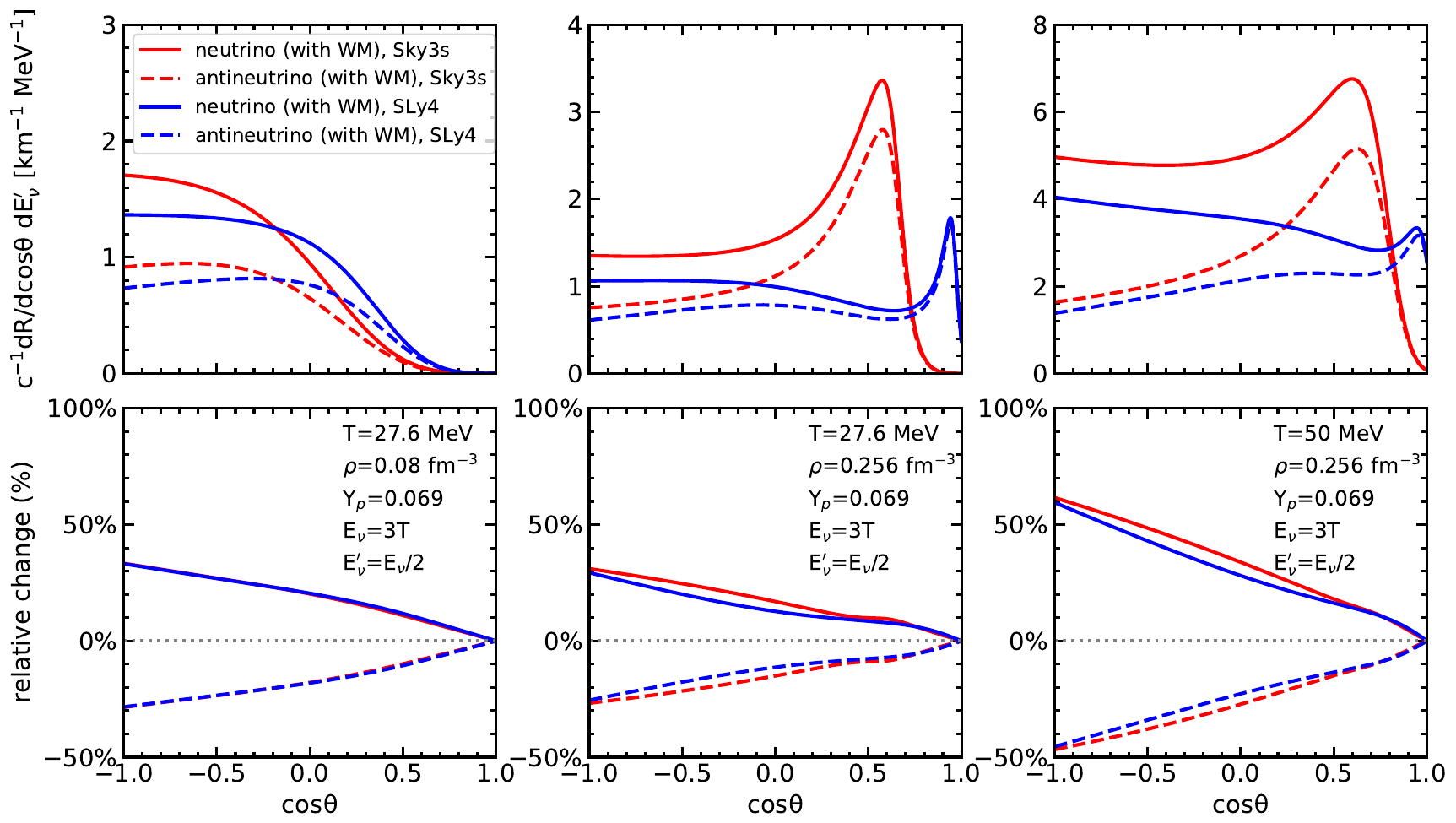} 
\caption{Top panels: double-differential neutrino (solid lines)
    and antineutrino (dashed lines) scattering rates $dR/d\cos\theta
  dE'_{\nu}$, computed with Sky3s (red) and SLy4 (blue)
    interactions for conditions (a), (b), and (c) given in Table
  \ref{table:conditions}, as functions of $\cos \theta$ with initial
  neutrino energy $E_{\nu}=3T$ and final neutrino energy
  $E'_{\nu}=\frac{E_{\nu}}{2}$. Bottom panels: the corresponding
  changes compared to calculations
    without WM corrections.}
\label{fig:R-cos-scattering}
\end{center}
\end{figure*}
\begin{figure*}[tb!]
\begin{center}
\includegraphics[scale=0.6]{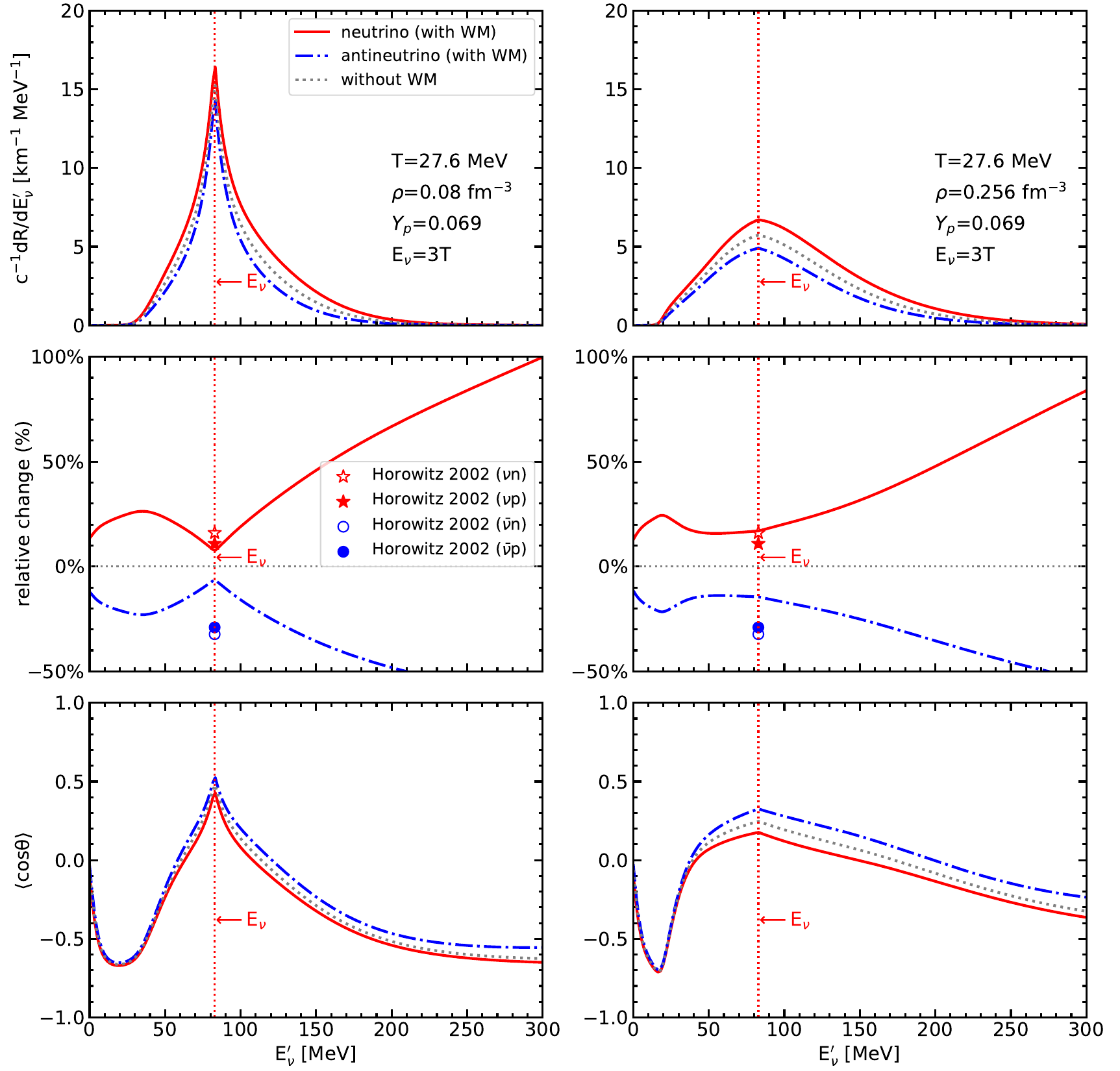} 
\caption{Differential neutrino (red solid lines) and
  antineutrino (blue dash-dotted lines) scattering rates
  $dR/dE'_\nu$ (top panels) and their relative change due to the WM
  correction (middle panels) as functions of the final neutrino energy
  $E'_{\nu}$ with the fixed initial neutrino energy $E_{\nu}=3T$, for
  conditions (a) and (b) in Table \ref{table:conditions}. The
  symbols represent the correction factors given by Horowitz in
  Ref. \cite{Horowitz2002}. Bottom panels: the corresponding evolution
  of the $\langle\cos\theta\rangle$.}
\label{fig:R-E1-costheta-E1-scattering}
\end{center}
\end{figure*}
\begin{figure*}[tb!]
\begin{center}
  \includegraphics[scale=0.6]{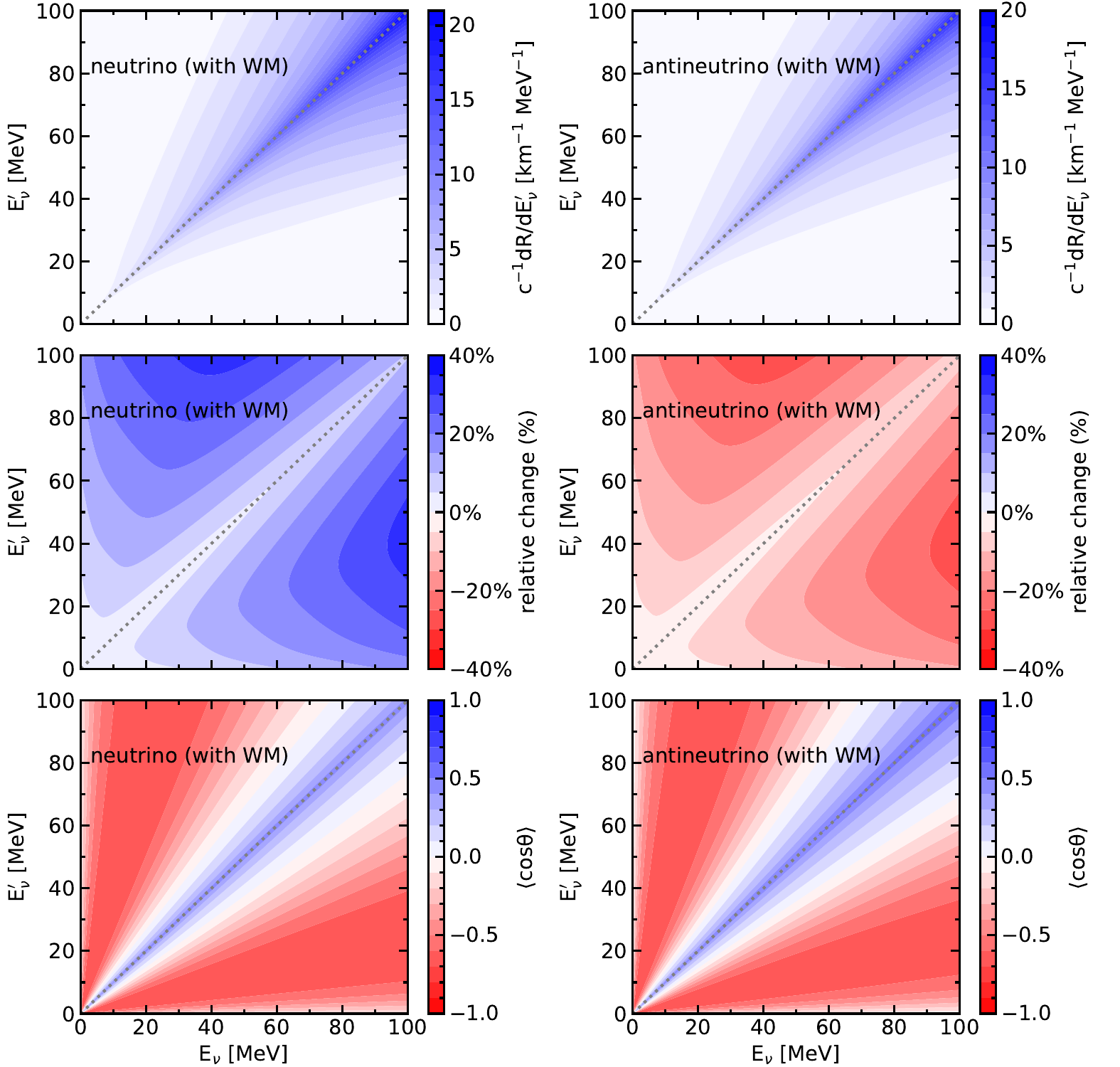}
\caption{Differential neutrino and antineutrino scattering rates (top
  panels), the corresponding relative changes of scattering rates due
  to WM (middle panels), and $\langle\cos\theta\rangle$ (bottom
  panels), computed at condition (a), i.e., $T=27.6$ MeV,
  $\rho=0.08\,\, \fm^{-3}$, $Y_p=0.069$.}
\label{fig:R0-cos-E1}
\end{center}
\end{figure*}

\section{Neutrino production and absorption}\label{sec:production-absorption}
As mentioned in the Introduction, many Skyrme interactions have
pathological or unrealistic properties, and the alternative extended
Skyrme functionals Sky3s and Sky4s were constructed recently to solve
these problems \cite{Duan2024,Duan2025}. In this section, we use the
Skyrme functional Sky3s (Sky3s and Sky4s give similar results under
the chosen conditions \cite{Duan2026}) to compute the modified
neutrino production and absorption rates and direct Urca neutrino
emissivities, taking into account WM corrections, and quantify the
modifications relative to those obtained without WM.

To study the WM effects on neutrino rates in hot nuclear matter, we
choose temperature $T=27.6$ MeV, baryon number density $\rho=0.256$
fm$^{-3}$, and proton fraction $Y_p=0.069$, from
Ref. \cite{Pascal2022}. To see whether the WM effects remain relevant
at subsaturation density, we also change the baryon number density to
$\rho=0.08$ fm$^{-3}$ but keep the temperature and proton fraction
fixed. Muon neutrinos and antineutrinos may be produced under these
conditions, but Ref. \cite{Pascal2022} did not consider
muons. Therefore, we can only roughly estimate the electron and muon
fractions $Y_e$ and $Y_{\mu}$ by requesting $\mu_{\mu}=\mu_{e}$ under
the constraint $Y_e+Y_{\mu}=Y_p$. For convenience, we label the low-
and high-density conditions as (a) and (b), respectively, and list the
parameters in Table \ref{table:conditions}.
  
\begin{table}[b!]
    \caption{Conditions used in the calculations of this paper.
      \label{table:conditions}}
    \begin{ruledtabular}
      \begin{tabular}{llllll}
        label & $\rho$     & $T$   & $Y_p$ & $Y_e$ & $Y_{\mu}$\\
              & (fm$^{-3}$) & (MeV) &       &       &\\
        \hline
        (a)   & 0.08       & 27.6  & 0.069 & 0.054 & 0.015 \\
        (b)   & 0.256      & 27.6  & 0.069 & 0.048 & 0.021 \\
        (c)   & 0.256      & 50    & 0.069 &       &
      \end{tabular}
    \end{ruledtabular}
\end{table}

The $F_\nu$ factor in Eq.
\eqref{eq:differential rates for all processes} implies that the
absorption and emission rates can be written as
$dR_{\abso,\nu}/d^3p d^3x = n^{\nu}_{\pv} \kappa_{\abso,\nu}(E)$ and
$dR_{\emis,\nu}/d^3p d^3x = (1-n^{\nu}_{\pv}) \kappa_{\emis,\nu}(E)$,
respectively. As explained in \cite{Burrows2006,Oertel2020}, it is
enough to show the sum $\kappa_\nu = \kappa_{\abso,\nu} +
\kappa_{\emis,\nu}$, from which the absorption and emission opacities
can be obtained as
\begin{equation}
  \kappa_{\abso,\nu} = (1-f_{\text{eq},\nu})\kappa_\nu\,,\quad
  \kappa_{\emis,\nu} = f_{\text{eq},\nu}\kappa_\nu\,,
  \label{eq:kappa-ab-em}
\end{equation}
where $f_{\text{eq},\nu_f}(E) = 1/(e^{(E-\mu_f+\mu_p-\mu_n)/T}+1)$ for
neutrinos and $f_{\text{eq},\bar{\nu}_f}(E) =
1/(e^{(E+\mu_f+\mu_n-\mu_p)/T}+1)$ for antineutrinos.

We show in Fig. \ref{fig:opacity-relative-change} the (anti)neutrino
opacities with WM corrections as functions of neutrino energy
$E_{\nu}$ (top panels) and the corresponding relative change compared
to those without WM corrections (bottom panels). Left and right panels
show the results under the conditions (a) and (b), respectively. At
the first glance, we can see in the bottom panels that the neutrino
opacities (red and blue solid lines) are always enhanced after adding
the WM corrections. This is consistent with the findings in
Refs. \cite{Horowitz2002,Sugiura2022}. However, at very low energies,
WM corrections enhance also the antineutrino opacities computed with
Sky3s (red and blue dashed lines). This is qualitatively different
from the conclusion of Refs. \cite{Horowitz2002,Sugiura2022} that the
WM corrections always decrease the antineutrino opacity.

For a more quantitative comparison, the orange solid and dashed lines
in the bottom panels of Fig. \ref{fig:opacity-relative-change}
represent the relative changes of neutrino absorption $\nu_e + n\to
e^- + p$ and antineutrino absorption $\bar{\nu}_e + p \to e^+ + n$
given by Horowitz in Ref. \cite{Horowitz2002}. These correction
factors have been used to add the WM corrections in supernova
simulations, such as Refs. \cite{Buras2006,Bollig2017,Ng2024}. They
are easy to use because they depend only on the neutrino energy and
not on the nuclear matter properties or interactions. However, we can
see from Fig. \ref{fig:opacity-relative-change} that they
underestimate the corrections for neutrinos at all energies. For
antineutrinos, they are close to ours at energies above 20 MeV under
condition (a) and between 12 and 50 MeV under condition (b). Below
these energies, the antineutrino opacities are strongly affected by
the neutron $\beta$ decay, while the correction factors of
Ref. \cite{Horowitz2002} were only derived for processes of the type
given in Eq. \eqref{eq:cc-process22}.

Furthermore, at small energies, the corrections of
Ref. \cite{Horowitz2002} are linear in energy, while we find a
positive offset. We find this offset already at the Hartree-Fock level
(without RPA). It is due to the motion of the initial nucleons and
their finite mass, which are considered higher-order effects in
Ref. \cite{Horowitz2002}. Notice that, once all processes of
Eqs. \eqref{eq:cc-process13} and \eqref{eq:cc-process22} are included,
neutrino and antineutrino opacities are equal at $E_\nu\to 0$. This
can be seen, e.g., from the fact that in Eq.
\eqref{eq:differential rates for all processes}, the sign $\Sign_{\nu}$
appears always together with a factor $E_{\nu}$. Hence, the WM
correction must have the same sign for neutrinos and antineutrinos at
low energies.

Besides, we can see that WM corrections are also important for muon
processes, and they are different from the electron ones because of
the large muon mass.

Now we turn to the direct Urca cooling in $\beta$ equilibrium. The
processes $e^{-} +p \rightarrow \nu_e + n$ and $n \rightarrow p +
e^{-} + \bar{\nu}_e$ are the most efficient cooling processes for
neutron stars. Here, we will not consider absorption processes and
assume that all neutrinos leave the star ($n^{\nu}_{\pv} = 0$). The
threshold density (at $T=0$) for direct Urca processes with Sky3s is
$\approx 0.47$ fm$^{-3}$ \cite{Duan2024}. At such high densities,
muons are also present (because $\mu_e > m_{\mu}$) and the direct Urca
processes $\mu^{-} + p \rightarrow \nu_{\mu} +n$ and $n \rightarrow p
+ \mu^{-} + \bar{\nu}_{\mu}$ can also occur. We can relate the
neutrino emissivities to the differential emission rates
\eqref{eq:differential rates for all processes} by the integral
\begin{equation}\label{eq:cooling-rates}
Q_{\nu} = \int dE_\nu 4 \pi E_{\nu}^3 \frac{dR_{\emis,\nu}}{d^3
  p_{\nu} d^3 x}.
\end{equation}
At sufficiently low temperature, $Q_\nu$ is proportional to $T^6$
\cite{Lattimer1991}, so that the coefficient $Q_\nu/T^6$ is
temperature independent.

We show in Fig. \ref{fig:direct-Urca} the direct Urca emissivities
(total neutrino and antineutrino emissivities) of neutron star matter
under the $\beta$ equilibrium conditions $\mu_n-\mu_p = \mu_e =
\mu_\mu$ (valid at low temperature) and $Y_p=Y_e+Y_{\mu}$. The results
were computed at $T=1$ MeV, and we checked that the emissivity was to
a good precision proportional to $T^6$. The red and blue curves
represent the results computed with and without WM contributions,
respectively. We limit ourselves to densities above 0.55 $\fm^{-3}$
because we found that at lower densities the equilibrium between muon
capture and muon production is very sensitive to temperature effects
even at $T=1$ MeV. As shown in Fig. \ref{fig:direct-Urca}, the
inclusion of WM enhances the direct Urca emissivities (red lines)
compared to the results obtained without this correction (blue
lines). The effect increases with density and it is stronger for
electron neutrinos and antineutrinos (top panel) than for muon ones
(bottom panel).

In our RPA calculation, WM can approximately double the electron
neutrino and antineutrino emissivities at high density. However, one
must be careful when comparing our results with those of
Refs. \cite{Leinson2002,Leinson2002b}, which find an increase of
direct Urca emissivities by at most $40-50\%$. The results of
Refs. \cite{Leinson2002,Leinson2002b} are computed within a
relativistic approach. Therefore, the numbers $40-50\%$ refer only to
the increase due to the anomalous magnetic moments ($\propto g_M$)
compared to results containing already the effect of the Dirac
magnetic moments ($\propto g_V$). If we make a similar comparison, we
get an enhancement of $46-67\%$ for $\nu_e$ and $\bar{\nu}_e$ and
$36-60\%$ for $\nu_{\mu}$ and $\bar{\nu}_{\mu}$, instead of the
numbers shown in Fig. \ref{fig:direct-Urca}.

Note that Leinson's results in Refs. \cite{Leinson2002,Leinson2002b}
are computed at baryon number density from about $2\rho_0$ to
$8\rho_0$, while we compute from about $3.4 \rho_0$ to
$5\rho_0$. Especially, within Leinson's relativistic approach, matter
contains also hyperons, such as $\Sigma^{-}$, $\Sigma^{0}$, and
$\Lambda$, apart from $npe\mu$. Therefore, a more detailed comparison
between the emissivities shown in our Fig. \ref{fig:direct-Urca} and
those in Refs. \cite{Leinson2002,Leinson2002b} would not make much
sense.
\section{Neutrino scattering}\label{sec:scattering}
Neutrino scattering also plays a crucial role in neutrino transport
which is needed in supernova simulations. In this section, we show the
scattering results.

We show in Fig. \ref{fig:R-cos-scattering} the double-differential
neutrino scattering rates and the corresponding relative changes after
adding WM corrections, as a function of $\cos \theta$ for the example
of a neutrino with initial energy $E_{\nu}=3T$ and final energy
$E_{\nu}^{\prime}=\frac{E_{\nu}}{2}$. Left, middle, and right panels
show the results under the conditions (a), (b), (c) listed in Table
\ref{table:conditions}, respectively.

In Ref. \cite{Duan2023}, we computed the scattering rates (without WM
corrections) with common Skyrme interactions, such as SLy4. We found
that these interactions predict far too low nucleon effective masses
$m^*_{n,p}$ at astrophysically relevant densities, leading, e.g., to
neutron Fermi velocities $v_{F,n} = k_{F,n}/m^*_n$ that exceed the
speed of light ($k_{F,n}$ denotes the neutron Fermi momentum). This
was one of our motivations to develop the Sky3s interaction in
Refs. \cite{Duan2024,Duan2025}. Therefore, let us compare in
Fig. \ref{fig:R-cos-scattering} scattering rates computed with SLy4
and Sky3s interactions under the same conditions. We can immediately
see that, with SLy4 (blue lines), the scattering rate does not go to
zero for vanishing scattering angle ($\cos\theta\to 1$, corresponding
to $\omega=q$) at $\rho = 0.256$ fm$^{-3}$ (middle and right
panels). Although $v_{F,n}$ is still slightly below the speed of light
at this density, the finite-temperature tail of the nucleon Fermi
distribution and the existence of a zero-sound mode in the response
function at $\omega > v_F q$ allow for timelike four-momentum
transfers (i.e., $\omega > q)$ in the SLy4 case. With the Sky3s
interaction (red lines), this problem does not appear at $T=27.6$ MeV
(middle panel), since the Fermi velocity and the speed of the
zero-sound mode are clearly below the speed of light. Only at the
highest temperature ($T=50$ MeV, right upper panel) there is a very
small rate at $\cos\theta=1$ due to the exponential tail of the
nucleon Fermi distribution.

Let us now come back to the discussion of the WM effects. Notice that,
without WM, the neutrino (solid lines) and antineutrino (dashed lines)
scattering rates would be equal. It has already been stressed in
previous studies, such as Refs. \cite{Horowitz2002,Sugiura2022}, that
WM corrections increase (decrease) neutrino (antineutrino) scattering
rates. This can be inferred from the last term of
Eq. \eqref{eq:scattering rate-sum}. The bottom panels in
Fig. \ref{fig:R-cos-scattering} show the relative change of neutrino
and antineutrino scattering rates compared to those without WM. We can
see that the increase of neutrino scattering rates is slightly larger
than the decrease of the antineutrino scattering rates. This is a
consequence of the term quadratic in $\tilde{C}_M$ in
Eq. \eqref{eq:scattering rate-sum}. The WM corrections get larger and
larger with increasing scattering angle (decreasing
$\cos\theta$). This is a consequence of the fact that the WM term in
the nucleon weak current Eq.
\eqref{eq:nucleon weak neutral current-approx-WM} is proportional to
the momentum transfer $\qv$, analogously to the explanation below
Eq. \eqref{eq:jCC} for the charged-current case. The factor
$(1-\cos\theta)$ in the last term of \eqref{eq:scattering rate-sum}
arises from the identity
$(E'_{\nu}\pv_{\nu}-E_{\nu}\pv'_{\nu})\cdot\qv =
E_{\nu}E'_{\nu}(E_{\nu}+E'_{\nu})(1-\cos\theta)$. Besides, as the
typical initial and final neutrino energies increase with temperature,
the WM corrections are expected to be stronger at higher temperature
(bottom right panel).

Using Eqs. (45) and (46) of Ref. \cite{Duan2023}, we can compute
differential scattering rates and the corresponding average cosine of
the scattering angle, $\langle\cos\theta\rangle$. The latter quantity
is particularly useful if the angle dependence of the neutrino
distribution function $n^{\nu}_{\pv}$ is expanded in Legendre
polynomials when solving the neutrino transport
\cite{Bruenn1985}. Figure \ref{fig:R-E1-costheta-E1-scattering} shows
the differential neutrino and antineutrino scattering rates (top
panels) as functions of the final neutrino energy $E_{\nu}^{\prime}$
for a fixed initial neutrino energy $E_{\nu}=3T$, as well as the
corresponding evolution of $\langle\cos\theta\rangle$ (bottom
panels). The relative changes of the differential scattering rates
compared to the calculation without WM are shown in the middle
panels. We choose again the combinations (a) and (b) of temperature,
density, and proton fraction listed in Table
\ref{table:conditions}. For comparison, the points marked as pentagram
and circle show the correction factors from Ref. \cite{Horowitz2002}
for elastic scattering ($E'_{\nu}=E_{\nu}$). We can see from
Fig. \ref{fig:R-E1-costheta-E1-scattering} that WM corrections
increase (decrease) both differential neutrino (antineutrino)
scattering rates and the corresponding average scattering angle (note
that increasing scattering angle corresponds to decreasing
$\cos\theta$), as one could have expected from
Fig. \ref{fig:R-cos-scattering}. From the middle panels of
Fig. \ref{fig:R-E1-costheta-E1-scattering}, we can also see that the
WM corrections for elastic scattering become larger with increasing
density in our case, whereas Horowitz's correction factors are the
same at different densities because they depend only on the energy.

Finally, we show in Fig. \ref{fig:R0-cos-E1} the differential
scattering rates (top panels) and $\langle\cos\theta\rangle$ (bottom
panels) as functions of initial and final energies for condition (a)
in Table \ref{table:conditions}. The corresponding relative changes of
scattering rates compared to the results that do not include WM are
also shown in the middle panels. Obviously, WM corrections increase
(decrease) differential neutrino (antineutrino) scattering rates and
the corresponding scattering angle for all combinations of initial and
final energies, but the corrections are different for different
initial and final energies.

\section{Conclusion}\label{sec:conclusion}
This work aims to study the effects of weak magnetism (WM) corrections
on (anti)neutrino weak processes in hot nuclear matter in the
framework of a nonrelativistic RPA calculation. We have incorporated
the WM term into the nucleon weak charged and neutral currents and
obtained the modified expressions for (anti)neutrino production,
absorption, and scattering rates. We perform our computations using
the recent nonrelativistic Skyrme functional Sky3s to avoid some
pathologies present in many other functionals. Because of the
interference between the WM term and the leading-order axial-vector
term, it is not possible to simply add the WM contribution on top of
the other contributions. We compare results that include WM
contributions with those obtained by considering only the
leading-order (in the nonrelativistic expansion) vector and
axial-vector currents.

For the charged-current processes, our results confirm that the weak
magnetism increases the electron and muon neutrino opacities and
decreases the antineutrino opacities except at low energies. However,
at low energies, it increases both the neutrino and antineutrino
opacities. Quantitatively, the corrections that we find in our
calculation can be only very roughly described by correction factors
obtained in Ref. \cite{Horowitz2002}, which depend only on the
(anti-)neutrino energy, underestimate the enhancement of the neutrino
opacities, and do not predict any enhancement of antineutrino
opacities at low energies. Because of the enhancement of neutrino- and
antineutrino production at low energies, the inclusion of WM
corrections result in a systematic increase of the cooling efficiency
whenever the direct Urca process can take place, in qualitative
agreement with Refs. \cite{Leinson2002,Leinson2002b}. Especially, WM
can nearly double the neutrino emissivities via electron processes at
high density.

Similarly to the charged-current case, our results confirm that WM
corrections also increase neutrino and decrease antineutrino
scattering rates. WM corrections and thus the difference between the
neutrino and antineutrino scattering rates become larger with
increasing scattering angle and neutrino energy, but they depend also
on other aspects, such as final energy and density, and so on, which
cannot easily be included in correction factors having a simple
analytical form. Even if WM does not lead to dramatic changes in the
scattering rates, it can affect the evolution of the composition of
matter during the supernova since it generates an asymmetry between
neutrinos and antineutrinos.

In conclusion, WM is an important ingredient in precise calculations
of neutrino transport in astrophysical simulations. The modification
of the formulas \eqref{eq:differential rates for all processes} and
\eqref{eq:scattering rate-sum} for (anti)neutrino production,
absorption, and scattering rates do not require any additional effort
compared to those without WM, since the numerical difficulties reside
in the calculation of the nuclear matter response functions and the
integrations over their imaginary parts, while WM adds only some new
prefactors to the integrand.

\begin{acknowledgments}
  The authors thank H.-T. Janka for useful discussions.
\end{acknowledgments}

\section*{Data availability}
The results shown in Figs. 1--5 are available from Ref.
\cite{Duan2026-wm-data}.

\bibliography{references}

\end{document}